\documentclass[12pt]{article}

\usepackage{graphicx}

\usepackage{epstopdf}
\begin{document}

\title{{\bf The Standard Model with Modified Asymmetric Potential}}
\author{
{Vladimir I. Kruglov}\\
{\em Physics Department, The University of Auckland,}\\ {\em Private Bag
92019, Auckland, New Zealand}\\
 e-mail: vik@phy.auckland.ac.nz
}
\date{}
\maketitle

\begin{abstract}
We consider a version of special relativity assuming that the metric in inertial frames is conformally pseudoeuclidean and depends on some scalar field with zero vacuum average. Applying this modified special relativity to the theory of electroweak interactions, and assuming that the introduced scalar field is proportional to the Higgs field, we show there is full consistency between the minimal Standard Model and our approach, except that, in our model, the potential in the Higgs sector is asymmetric and has two distinct vacuum states.  It is found that, in the new modified model the mass of the Higgs particle is half that in the minimal Standard Model.
\end{abstract}

PACS numbers {03.30.+p, 12.15.-y, 12.60.Fr}

\section{Introduction}

Investigating the consistency of quantum mechanics in extreme situations, such as black holes, has led some authors \cite{Sus, Gar} to suggest a radical revision of the kinematics of systems with very high energy. They proposed \cite{Sus, Gar} a modification of special relativity in which the Planck energy $E_P$ or length $l_P$ joins the speed of light $c$ as an invariant. 

This concept based on the assumption that the Lorentz invariance is only an approximate symmetry has been developed in \cite{Am1, Am2, Bru} on the basis of $\kappa$-Poincar\'e Hopf algebras \cite{Luk, Maj, Lu}. Another modification of special relativity \cite{Joa} is based on a nonlinear generalization of the action of the Lorentz group on momentum space, and can be interpreted as the Fock-Lorentz symmetry \cite{Fock} applied to momentum space. There have been also numerous claims that Lorentz symmetry breaking could be observable  in the near future or may even already have been observed \cite{El2, Gam, Adu, Alf}.

We propose in this Letter a different generalization of special relativity which supposes that the Minkowski spacetime structure begins to perturb at energies much lower then the Planck energy $E_P$.  In particular, our model predicts that the deviations from special relativity start to become significant at energies comparable to the mass of the Higgs particle which is the same order as the vacuum expectation value $v=247$~GeV of the Higgs field $\phi(x)$. Because our model of spacetime is Lorentz invariant, the deviations from special relativity have mainly dynamical character when gravitational effects are negligible. 

The modification is connected with the Weinberg-Salam theory of electroweak interactions \cite{Wei,  SW} and is based on the assumption that there exist some fine-grained perturbations of the vacuum state described by a scalar field $\chi(x)$. Moreover, we assume that these perturbations take place in any inertial frame and the vacuum average of the scalar field $\chi(x)$ is zero.  

Following Einstein's special relativity, we assume that all physical processes are equal in any inertial frame. We also require that the metric of the perturbed space in any inertial frame is given by the fundamental quadratic form (see Eq.(\ref{1})) with new interval $d\tilde{s}=\left(1+\chi(x)\right)ds$, where $ds$ is the interval in the Minkowski pseudoeuclidean space. Using this metric we show that the conservation laws are fulfilled for average $4$-vector momentum and $4$-tensor angular momentum because the vacuum average operator flattens the spacetime perturbations: $\langle\chi(x)\rangle=0$. 

 In fact, the generalized special relativity yields a procedure for obtaining the mass terms in the Lagrangian of the minimal Standard Model (SM) depending on the Higgs field. This procedure is alternative to the Higgs mechanism \cite{Hig} which allows the introduction of mass terms to the Yang-Mills fields \cite{YM, GG} without violation of the renormalization principle.  
 
It is remarkable that the approach developed in this paper leads to the same mass terms in the Lagrangian as the minimal SM but to a different potential for the Higgs field. In our modified model the  potential $V_m(\phi)$ is asymmetric and has two distinct vacuum states at $\phi=0$ and $\phi=v$ in contrast to the symmetric potential $V(\phi)$ of the minimal SM with two equivalent vacuum states at $\phi=-v$ and $\phi=v$. Moreover, in the model with asymmetric  potential the mass of the Higgs particle is half that in the minimal SM.
 To complete the introduction, we remark that the quantization of the Higgs field leads to a version of quantum geometry.

\section{Generalized special relativity}

We start with a postulate of the conformally pseudoeuclidean metric form in generalized special relativity:

$\circ$~{\it The fundamental quadratic form of the interval $d\tilde{s}$ in any inertial frame is forminvariant and given as}~:
\begin{equation}
d\tilde{s}^2=\left(1+\chi(x)\right)^2\left((dx^0)^2-(dx^1)^2-(dx^2)^2-(dx^3)^2 \right),
\label{1}
\end{equation}
{\it where the vacuum average of the scalar field $\chi(x)$ is zero} ($\langle\chi(x)\rangle=0$).

This fundamental quadratic form is invariant under the Lorentz transformations and can be written also via the metric tensor $g_{\alpha\beta}(x)$~:
\begin{equation}
d\tilde{s}^2=g_{\alpha\beta}(x)dx^\alpha dx^\beta,~~~g_{\alpha\beta}(x)=\left(1+\chi(x)\right)^2\eta_{\alpha\beta},
\label{2}
\end{equation}
 where $\eta_{\alpha\beta}$ is the diagonal Minkowski tensor. Note that the transformations of the conformal group \cite{DFN} are defined as $g_{\alpha\beta}(x)\rightarrow \lambda(x)g_{\alpha\beta}(x)$ with $\lambda(x)>0$. This conformal group is isomorphic to ${\rm O}(4,2)$ and locally isomorphic to ${\rm SU}(2,2)$. 
 
We begin with relativistic classical mechanics based on the generalized Poincar\'e action for a free particle~:
\begin{equation}
\mathcal{S}=-mc\int_a^bd\tilde{s}=\int_{t_1}^{t_2}\mathcal{L}(x)dt,~~~\mathcal{L}(x) = -mc\frac{d\tilde{s}}{dt},
\label{3}
\end{equation}
then the Lagrangian is
\begin{equation}
\mathcal{L}(x) = -mc\sqrt{g_{\alpha\beta}(x)\frac{dx^\alpha}{dt}\frac{dx^\beta}{dt}}=-mc^2\sqrt{1-\beta^2}\left(1+\chi(x)\right),
\label{4}
\end{equation}
where $\beta={\rm v}/c$, and ${\rm v}$ is the velocity of the particle.

Hence the energy and momentum formally defined by the Lagrangian (\ref{4}) depend on the scalar field $\chi(x)$ as~:
\begin{equation}
{\rm E}(x)= {\rm {\bf v}}\frac{\partial \mathcal{L}(x)}{\partial {\rm {\bf v}}} -  \mathcal{L}(x)=\frac{mc^2}{\sqrt{1-\beta^2}}\left(1+\chi(x)\right), 
\label{5}
\end{equation}

\begin{equation}
{\rm {\bf P}}(x)=\frac{\partial \mathcal{L}(x)}{\partial {\rm {\bf v}}}=\frac{m{\rm {\bf v}}}{\sqrt{1-\beta^2}}\left(1+\chi(x)\right).
\label{6}
\end{equation}
These formally defined the energy and the momentum are not the constants of motion because the interval $d\tilde{s}$ is not invariant under translation group. However the vacuum averaged interval is invariant under translation group: $\langle d\tilde{s}\rangle=ds=constant$ because $\langle\chi(x)\rangle=0$, and
hence average energy ${\rm E}=\langle{\rm E}(x)\rangle$ and momentum ${\bf P}=\langle{\rm {\bf P}}(x)\rangle$ are constants of motion as directly follows from Eqs.(\ref{5},{6})~: 
\begin{equation}
{\rm E}=\langle{\rm E}(x)\rangle=\frac{mc^2}{\sqrt{1-\beta^2}},~~~{\bf P}=\langle{\rm {\bf P}}(x)\rangle=\frac{m{\rm {\bf v}}}{\sqrt{1-\beta^2}}.
\label{7}
\end{equation}
Thus the conservation laws given by Eq.(\ref{7}) arise because the averaging operator reduces inhomogeneous space given by the quadratic form Eq.(\ref{1}) to the homogeneous Minkowski space which is invariant under four-parametric translation group. We also introduce new (own) coordinates $\tilde{x}^\alpha$ which do not define any inertial frame but which transform the fundamental quadratic form into~:
\begin{equation}
d\tilde{s}^2=\eta_{\alpha\beta}d\tilde{x}^\alpha d\tilde{x}^\beta= d\tilde{x}^\alpha d\tilde{x}_\alpha,~~~ d\tilde{x}^\alpha=\left(1+\chi(x)\right)dx^\alpha,
\label{8}
\end{equation}
where $d\tilde{x}_\alpha=\eta_{\alpha\beta}d\tilde{x}^\beta$.
Using the new coordinates $\tilde{x}^\alpha$ one can represent the momentum 4-vector in the form~:
\begin{equation}
{\rm P}^\alpha(x) \equiv mc\frac{d\tilde{x}^\alpha}{ds}=\left(\frac{{\rm E}(x)}{c},{\rm {\bf P}}(x)\right),~~~{\rm P}_\alpha(x)\equiv mc\frac{d\tilde{x}_\alpha}{ds}=\left(\frac{{\rm E}(x)}{c},-{\rm {\bf P}}(x)\right),
\label{9}
\end{equation}
where the covariant 4-vector is defined here as ${\rm P}_\alpha(x) \equiv \eta_{\alpha\beta}{\rm P}^\alpha(x)$. Hence {\it though the metric tenser in coordinate space is given as $g_{\alpha\beta}(x)=\left(1+\chi(x)\right)^2\eta_{\alpha\beta}$, the corresponding metric tensor in the momentum space is $\eta_{\alpha\beta}$}. That the metric tensor in the momentum space is $\eta_{\alpha\beta}$ can also be derived by assuming   
that the contravariant ${\rm P}^\alpha=\langle{\rm P}^\alpha(x)\rangle$ and covariant ${\rm P}_\alpha=\langle{\rm P}_\alpha(x)\rangle$ momentum 4-vectors coincide with the corresponding momentum 4-vectors in special relativity. The Eqs.(\ref{9}) yield the dispersion equation as~: 
\begin{equation}
\eta^{\alpha\beta}{\rm P}_\alpha(x){\rm P}_\beta(x)= m^2 c^2\left(1+\chi(x)\right)^2.
\label{10}
\end{equation}

Considering the integral in Eq.(\ref{3}), with the upper limit a function of the point $x$ ($b=b(x)$) and a fixed lower limit $a$, one can derive the equation ${\rm P}_\alpha(x)=-\partial \mathcal{S}/\partial x^\alpha$. Using this expression and Eq.(\ref{10}) we find the equation of a geodesic line in Hamilton-Jacobi form as~:  
\begin{equation}
\eta^{\alpha\beta}\frac{\partial \mathcal{S}}{\partial x^\alpha}\frac{\partial \mathcal{S}}{\partial x^\beta} = m^2 c^2\left(1+\chi(x)\right)^2.
\label{11}
\end{equation}

In a field theory based on the fundamental quadratic form Eq.(\ref{1}) the conservation laws for the $4$-vector momentum ${\rm P}^\alpha$ and $4$-tensor angular momentum ${\rm M}^{\alpha\beta}$ can be written in the standard form~:  

\begin{equation}
{\rm P}^\alpha=\frac{1}{c}\int \langle \mathcal{T}^{\alpha\beta} \rangle d\mathcal{F}_\beta,~~~{\rm M}^{\alpha\beta}=\frac{1}{c}\int (x^\alpha \langle\mathcal{T}^{\beta\lambda}\rangle-x^\beta\langle\mathcal{T}^{\alpha\lambda}\rangle) d\mathcal{F}_\lambda,
\label{12}
\end{equation}
where $\langle \mathcal{T}^{\alpha\beta} \rangle$ is the stress-energy tensor defined as the vacuum average of the corresponding operator $\mathcal{T}^{\alpha\beta}$ and satisfying the equations
\begin{equation}
\langle \mathcal{T}^{\alpha\beta}(x)  \rangle=\langle \mathcal{T}^{\beta\alpha}(x) \rangle,~~~       \partial\langle \mathcal{T}^{\alpha\beta}(x)\rangle/\partial x^\alpha=0.
\label{13}
\end{equation}
In this model 10 conservation laws arise because the vacuum averaging operator transforms the inhomogeneous space given by the fundamental quadratic form Eq.(\ref{1}) to the homogeneous Minkowski space which is invariant under four-parametric translation group and six-parametric rotation group.

\section{Standard Model and asymmetric potential}

In this section we consider application of the generalized special relativity to the minimal SM,
assuming that the scalar field $\chi(x)$ is proportional to the Higgs field $h(x)=\phi-v$. More precisely we postulate:

$\circ$~{\it The scalar field $\chi(x)$ is equal to the normalized Higgs field} $\chi(x)=h(x)/v$ {\it i.e. the field $\chi(x)$ is connected with the Higgs doublet $\Phi(x)$ as}~:

\begin{equation}
\Phi(x)=\frac{1}{\sqrt{2}}\left(\begin{array}{c}
0\\
\phi(x)\\
\end{array}\right)=\left(1+\chi(x)\right)\Phi_0,~~~\Phi_0=\frac{1}{\sqrt{2}}\left(\begin{array}{c}
0\\
v\\
\end{array}\right).
\label{14}
\end{equation}

In this Letter, we consider the minimal SM, which requires one Higgs field doublet and predicts a single neutral Higgs boson, and we follow the notation in \cite{Quark}.

From Eqs.(\ref{10},{11}) it follows that we can use the tensor $\eta_{\alpha\beta}$ to construct the invariants under the Lorentz group in Lagrangians, {\em provided} the masses of the particles are replaced as $m\rightarrow m(1+\chi(x))$ (see also Eqs.(\ref{5},{6})).

This leads to the following procedure for obtaining the mass terms in a Lagrangian:
 
$\circ$~{\it First step: Introduce mass terms into the Lagrangian for different particles   
according to the usual phenomenological procedure which breaks the gauge invariance of the Lagrangian}.

 $\circ$~{\it Second step: The mass of every particle in the Lagrangian should be replaced according to}~: 
\begin{equation}
m\rightarrow m\left(1+\frac{h(x)}{v}\right),~~~\chi(x)=\frac{h(x)}{v}.
\label{15}
\end{equation}

 Thus, according to this procedure, the electron's mass term can be introduced in the
{\it First step} as $\mathcal{L}_e(x)=-m_e\bar{e}e$, and becomes in the {\it Second step:} $\mathcal{L}_e(x)=-m_e\left(1+h(x)/v\right)\bar{e}e$. The mass terms for $\mu$ and $\tau$ leptons follow  by substituting $e\rightarrow \mu, \tau$ in the Lagrangian $\mathcal{L}_e(x)$. Hence we obtain exactly the same expressions as in the minimal SM. Our procedure also yields terms for the six quarks which are the same as those in the minimal SM~:
\begin{equation}
\mathcal{L}_{u,d}(x)=-\sum_i m_u^i\left(1+\frac{h(x)}{v}\right)\bar{u}_iu_i - \sum_i m_d^i\left(1+\frac{h(x)}{v}\right)\bar{d}_id_i.
\label{16}
\end{equation}

Similarly the mass terms for gauge bosons obtained from the Lagrangian 
$\mathcal{L}_{W,Z}=|[-i(g/2)\tau_aW_\mu^a-i(g'/2)B_\mu]\Phi(x)|^2$ via diagonalization \cite{Weinberg} contain the factor $(1+h(x)/v)^2$~: 
\begin{equation}
\mathcal{L}_{W,Z}=M_W^2\left(1+\frac{h(x)}{v}\right)^2W_\mu^+W^{-\mu} + \frac{1}{2}M_Z^2\left(1+\frac{h(x)}{v}\right)^2Z_\mu Z^\mu,
\label{17}
\end{equation}
where $M_W=|v|g/2$, $M_Z=|v|\sqrt{g^2+g'^2}/2$, $g/g'={\rm cot}(\theta_W)$, and
$W^{\pm}=(W^1\pm W^2)/\sqrt{2}$. The terms for $W$ and $Z$ particles in Eq.(\ref{17}) follow automatically from the gauge invariance of the Lagrangian and moreover have the form consistent with the the above procedure.
Considering the Higgs sector, one can use the proposed procedure to find the potential $U_m(h)$ of the field $h(x)=\phi(x)-v$.

 {\it First step:} $\mathcal{L}_\Phi=\frac{1}{2}(\partial_\mu h)(\partial^\mu h)-\frac{1}{2}\mu_h^2h^2-\mathcal{E}_0$.  {\it Second step:} $\mathcal{L}_\Phi=\frac{1}{2}(\partial_\mu h)(\partial^\mu h)-U_m(h)$, where the potential $U_m(h)$ has the form
\begin{equation}
U_m(h)=\frac{1}{2}\mu_h^2\left(1+\frac{h(x)}{v}\right)^2h(x)^2+\mathcal{E}_0.
\label{18}
\end{equation}
Here $\partial_\mu h=\eta_{\mu\nu}\partial^\nu h$ and $\mathcal{E}_0$ is an arbitrary constant. Using the scalar field  $\phi(x)$ one can rewrite Eq.(\ref{18}) as $\mathcal{L}_\Phi=\frac{1}{2}(\partial_\mu \phi)(\partial^\mu \phi)-V_m(\phi)$, where the potential $V_m(\phi)$ is
\begin{equation}
V_m(\phi)=\frac{\mu_h^2}{2v^2} \phi^2(\phi-v)^2+\mathcal{E}_0.
\label{19}
\end{equation}
This modified asymmetric potential ($V_m(\phi)\neq V_m(-\phi)$) has two distinct vacuum states at $\phi=0$ and $\phi=v$ in contrast to the potential of the minimal SM which has two equivalent vacuum states at $\phi=\pm v$.
The potential $V_m(\phi)$ is symmetric about $\phi=v/2$ and has the form $V_m'(\phi')=\phi'^2(\phi'-1)^2$ (see Fig.1) for dimensionless variables $V_m'=2V_m/(v\mu_h)^2$ and $\phi'=\phi/v=1+\chi$ where we assume $\mathcal{E}_0=0$. 

\begin{figure}
\includegraphics[width=4in]{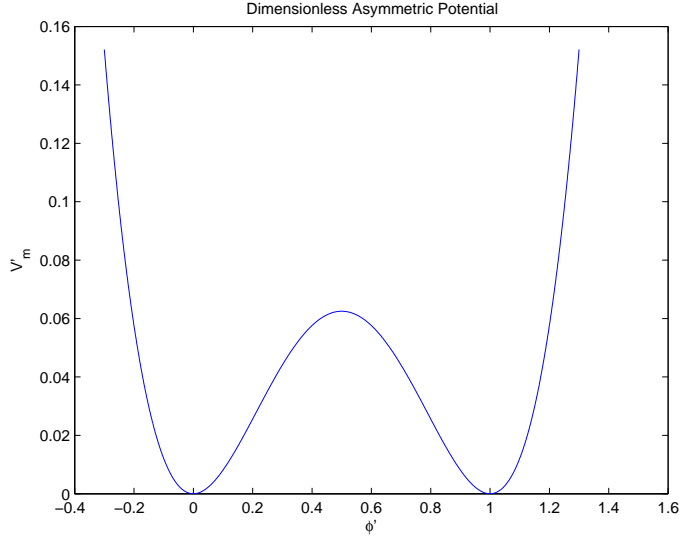}
\caption{The dimensionless asymmetric potential $V'_m(\phi')$ is symmetric about $\phi'=1/2$. Minima occur at $\phi'=0$ and $\phi'=1$, and these correspond to the two vacuum states at $\phi=0$ and $\phi=v$.}
\label{fig:1}
\end{figure}

Note that, in the new model with asymmetric potential $V_m(\phi)$ given by Eq.(\ref{19}), the mass $\mu_h$ of the Higgs particle  is only half the mass $m_h$ of the Higgs particle in the minimal SM with symmetric potential $V(\phi)$ (i.e. $\mu_h=m_h/2$).
To verify this statement we write the symmetrc potential $V(\phi)$ of the minimal SM as
$V(\phi)=m_h^2/(8v^2) (\phi-v_{-})^2(\phi-v_{+})^2+\mathcal{E}_0$,
where $v_{\pm}=\pm v$. 
Next we treat $v_{-}$ as a variable parameter. Hence, for the field $h(x)=\phi(x)-v$, the potential with the variable parameter $v_{-}$ is~: 
\begin{equation}
U(h)=\frac{\mu_h^2}{2(v_{+}-v_{-})^2} (v_{+}-v_{-}+h)^2h^2+\mathcal{E}_0,
\label{20}
\end{equation}
where we have introduced the mass $\mu_h$ of the Higgs particle depending on the parameter  $v_{-}$ as $\mu_h=m_h(v_{+}-v_{-})/(2v)$. In the limit $v_{-}\rightarrow 0$ we find from Eq.(\ref{20}) that $U(h)\rightarrow U_m(h)$ ($V(\phi)\rightarrow V_m(\phi)$) and $\mu_h\rightarrow m_h/2$. Note that, in this proof, $v_{+}$ is kept fixed because the value $\phi=v$ corresponds to a vacuum state in both models.
 We remark that the minimal SM and proposed modified model with asymmetric potential $V_m(\phi)$ are both invariant under the global transformation $\phi\rightarrow -\phi$ $\&$ $v\rightarrow -v$ as required for renormalization of the theory \cite{Hooft}.  
Let us finally estimate the barrier $\varepsilon_v$ (energy density) separating the two vacuum states at $\phi=0$ and $\phi=v$: $\varepsilon_v\equiv V_m(v/2)-V_m(v)=\mu_h^2v^2/32$.
Assuming $\hbar = c = 1$, $\mu_h\simeq 100$~GeV, and $v=247$~GeV, we find $\varepsilon_v \simeq 1.9\cdot10^7$~GeV$^4$ or $\varepsilon_v \simeq 2.5\cdot 10^{48}$~GeV/cm$^3$.
The energy density $\varepsilon_v$, as well as nonlinear (cubic and quartic) terms in the expansion of $U_m(h)$ given by Eq.(\ref{18}), represent differences between the Standard Model with symmetric and asymmetric potentials that can be recognized in future high energy experiments. The model presented here could also be extended to the minimal supersymmetric SM which introduces two Higgs field doublets \cite{Hab}.

Let us suppose that  all particles (except photons and neutrinos) in the Standard Model with asymmetric potential have masses $\sim v$. Hence in the limit $v\rightarrow 0$ the full Lagrangian  represents massless fields. Taking into account the tunnel effect, one can derive that some energy may exist in the form of excitations of the vacuum state $|0\rangle$ at $\phi=0$. 
Moreover, such tunneling effects would seem likely to occur in cosmic processes, such as the collapsing of stars, where the energy density may approach $\varepsilon_v\simeq 10^{48}$~GeV/cm$^3$.

\section{Conclusions}

In this Letter we have proposed a version of the special relativity assuming that the fundamental metric form in any inertial frame is conformally pseudoeuclidean and depends on the scalar field $\chi(x)$ which has zero vacuum average. Further we identify this scalar field with the normalized Higgs field:
$\chi(x)=h(x)/v$,  where $v$ is the vacuum expectation value. Because the new generalized interval $d\tilde{s}$ is not invariant under translation group, the energy and the momentum defined formally by an appropriate Lagrangian are not the constants of motion. However it is shown that corresponding vacuum expectation values can be consider as conserved energy and momentum. 
In this model, conservation of momentum, energy, and angular momentum arise because the vacuum averaging operator transforms the inhomogeneous space given by the fundamental quadratic form Eq.(\ref{1}) to the homogeneous Minkowski space which is invariant under four-parametric translation group and six-parametric rotation group. The application of the proposed version of the special relativity to the minimal SM leads to the 
procedure for obtaining the mass terms in a Lagrangian given by Eq.(\ref{15}). We show there is full consistency between the minimal SM and our approach, except the Higgs sector. The procedure (\ref{15}) yields the modified asymmetric potential $V_m(\phi)$ given by Eq.(\ref{19}) which has two distinct vacuum states at $\phi=0$ and $\phi=v$ in contrast to the potential of the minimal SM which has two equivalent vacuum states at $\phi=\pm v$. We show that in the new model with asymmetric potential $V_m(\phi)$ the mass $\mu_h$ of the Higgs particle is only half the mass $m_h$ of the Higgs particle in the minimal SM with usual symmetric potential.

Finally we remark that the introduced conformally pseudoeuclidean metric form is quadratic with respect to the Higgs doublet: $d\tilde{s}^2=(2/v^2)\Phi^{+}(x)\Phi(x)$ $\eta_{\alpha\beta}dx^\alpha dx^\beta$, and hence is invariant under gauge transformations. Thus the quantization of the Higgs field has led to a version of quantum geometry with the metric form: $d\hat{s}^2=\eta_{\alpha\beta}d\hat{x}^\alpha d\hat{x}^\beta$ where $d\hat{x}^\alpha=(1+\hat{\chi}(x))dx^\alpha$, which is compatible with the gauge field theory. Moreover,  $\langle d\hat{x}^\alpha\rangle=dx^\alpha$ because $\langle\hat{\chi}(x)\rangle=0$, and $\langle d\hat{s}\rangle=ds$ due to $d\hat{s}=(1+\hat{\chi}(x))ds$,
where $ds^2=\eta_{\alpha\beta}dx^\alpha dx^\beta$. Combining these formulae one can write :
\begin{equation}
ds^2=\eta_{\alpha\beta}d{x}^\alpha d{x}^\beta=\langle d\hat{s}\rangle^2=\eta_{\alpha\beta}\langle d\hat{x}^\alpha\rangle \langle d\hat{x}^\beta\rangle. 
\label{21}
\end{equation}

Hence, the generalized special relativity presented here takes into account quantum fluctuations of the spacetime coordinates $d\hat{x}^\alpha-dx^\alpha=\hat{\chi}(x)dx^\alpha$, and is compatible with Einstein's special relativity (the special relativity spacetime coordinates being defined as the vacuum average of the quantized spaces: $dx^\alpha=\langle d\hat{x}^\alpha\rangle$). 
The stronger assumption that coordinates do not commute was made a long time ago \cite{HS}. We also note that, in the  model of spacetime presented here, the kinematics of the particles is the same as in standard special relativity.
     
We suppose that the perturbation of the spacetime is the result of creation and annihilation of the virtual particle-antiparticle pairs in the empty space which we take into account by introducing the quantized Higgs field into pseudoeuclidean metric form.  Hence, empty space becomes densely populated with virtual particles, and empty space behaves as a dynamical medium described by quantized Higgs field.

\section*{Acknowledgements}

I am grateful to J. Harvey for his support and D. Wardle and B.S. Pavlov for helpful discussions.

%\section{References}

%\begin{references}

\end{document}